\documentclass[]{pasj01}       
\usepackage{bm}
\begin{document}        
\Received{$\langle${2017/May/31}$\rangle$}
\Accepted{$\langle${2017/Aug/27}$\rangle$}
       
\title{Theoretical modeling of Comptonized X-ray spectra of super-Eddington accretion flow:       
origin of hard excess in Ultraluminous X-Ray Sources}

\author{Takaaki \textsc{Kitaki},\altaffilmark{1}$^{,*}$       
  Shin \textsc{Mineshige},\altaffilmark{1}       
  Ken \textsc{Ohsuga},\altaffilmark{2}       
  and Tomohisa \textsc{Kawashima}\altaffilmark{2}       
}       
  \altaffiltext{1}{Department of Astronomy, Graduate School of Science, Kyoto University, Kitashirakawa-Oiwake-cho, Sakyo-ku, Kyoto 606-8502, Japan}       
  \altaffiltext{2}{National Astronomical Observatory of Japan, 2-21-1 Osawa, Mitaka-shi, Tokyo 181-8588, Japan}       
  \email{kitaki@kusasro.kyoto-u.ac.jp}       
       
       
\KeyWords{accretion, accretion disks --- radiation: dynamics --- stars: black holes} 
       
\maketitle       
       
\begin{abstract}      
X-ray continuum spectra of super-Eddington accretion flow are studied by means of   
Monte Carlo radiative transfer simulations based on the radiation hydrodynamic simulation data,   
in which both of thermal and bulk Compton scatterings are taken into account.      
We compare the calculated spectra of accretion flow around black holes      
with masses of $M_{\rm BH} = 10, 10^2, 10^3$, and $10^4 M_\odot$ for a fixed   
mass injection rate (from the computational boundary at $10^3 r_{\rm s}$) of  
$10^3 L_{\rm Edd}/c^2$ (with $r_{\rm s}$, $L_{\rm Edd}$, and $c$ being   
the Schwarzschild radius, the Eddington luminosity, and the speed of light, respectively).      
The soft X-ray spectra exhibit mass dependence in accordance with the standard-disk relation;     
the maximum surface temperature is scaled as $T \propto M_{\rm BH}^{-1/4}$.     
The spectra in the hard X-ray bands, by contrast, look quite similar     
among different models, if we normalize the radiation luminosity by $M_{\rm BH}$.     
This reflects that the hard component is created by thermal and bulk Compton scattering   
of soft photons originating from an accretion flow in the over-heated and/or funnel regions,   
the temperatures of which have no mass dependence.   
The hard X-ray spectra can be reproduced by a Wien spectrum with temperature of $T\sim 3$ keV      
accompanied by a hard excess at photon energy above several keV.  
The excess spectrum can be well fitted with a power law with a photon index of $\Gamma \sim 3$.  
This feature is in good agreement with that of the recent NuSTAR observations of ULX (Ultra-Luminous X-ray sources).   
\end{abstract}

\section{Introduction}       
It had been long believed that the classical Eddington limit can never be exceeded,       
as long as objects steadily shine by accreting environmental gas.        
Recently, however, there are growing evidences, which point the existence       
of astronomical objects shining at super-Eddington luminosities.       
Super-Eddington accretion flow is the flow, in which the accretion rate onto       
a central object exceeds the classical limit which gives rise to super-Eddington luminosities;     
that is,     
\begin{eqnarray}      
  \dot{M} > \dot{M}_{\rm Edd}/\eta &\equiv& L_{\rm Edd}/(\eta c^{2})\nonumber\\      
  &\simeq& 10^{19} ({\rm g~s}^{-1}) (\eta/0.1)^{-1}(M_{\rm BH}/10 M_\odot),       
\end{eqnarray}      
where $L_{\rm Edd}$ is the Eddington luminosity,        
$\eta (\sim 0.1)$ is the energy conversion efficiency,       
$M_{\rm BH}$ is the black hole mass, and $c$ is the speed of light.       
       
One of the best candidates of the super-Eddington accretors       
is Ultra-Luminous X-ray sources (ULXs), bright off-nuclear compact X-ray sources        
with luminosities of $10^{39}$--$10^{41}$ erg s$^{-1}$, found in nearby galaxies.       
There are two major scenarios explaining the high luminosities of ULXs:       
sub-Eddington accretion to intermediate-mass black holes (e.g., \cite{key3m}; \cite{key301}),     
and super-Eddington accretion to stellar mass black holes (e.g., \cite{key81}; \cite{key21}).       
      
The discovery of X-ray pulses in M82 X-2 by \citet{key0} was striking in this respect,        
since it demonstrates that the central star in M82 X-2 should be a neutron star        
and is thus less massive than $3 M_\odot$.       
Super-Eddington accretion should, hence, be required to explain its high luminosity        
over $10^{40}$ erg s$^{-1}$. The discovery such an object called a ULX pulsar provides       
a good support for the existence of super-Eddington accretion in the universe.       
Two other ULX pulsars were discovered afterward (NGC7793 P13, \cite{key01}; NGC5907 ULX, \cite{key02}).       
    
We should note that ULXs are not the only super-Eddington accretors but    
there exist other sites, where super-Eddington accretion is going on.    
Some of microquasars are suspected to have super-Eddington accretion flow     
(see, e.g., \cite{key00})    
and some quasars at high (cosmological) redshifts are another example.    
\citet{key31} discovered the quasars at the redshift, $z=7.085$. Its black hole mass is     
estimated to exceed $10^{9}M_{\odot}$.       
If this supermassive black hole had grown up from a population III remnant     
(a black hole with $M_{\rm BH}\sim 10$--$10^{3}M_{\odot}$),     
super-Eddington accretion flows is inevitable, since otherwise the growth time     
to produce the observed supermassive black hole exceeds the age of the universe.    
     
       
The inflow/outflow gas dynamics and emission properties of super-Eddington accretion        
have been extensively studied in this decade by means of multi-dimensional  
radiation hydrodynamic (RHD) simulations, being pioneered by \citet{key4}  
(see also \cite{key41}, \cite{key42} for radiation-MHD simulations).  
More recently, general relativistic (GR) RHD simulations have been performed       
(\authorcite{key52} \yearcite{key52}, \yearcite{key53}; \cite{key3000}; \cite{key61}).       
They all demonstrated that significant fraction of inflow material is blown away     
by strong radiation-pressure force in forms of optically Thomson-thick,      
moderately high-temperature ($\sim$ several keV) disk wind.  
\citet{key03} have reconfirmed such a structure by modified RHD simulations.      
The emergent spectra should thus suffer Comptonization effects by the wind.       
\citet{key2} were the first to calculate       
such Comptonized spectra of super-Eddington accretion flows and have shown       
that the theoretical spectra can well fit the typical ULX spectra in the X-ray band (\authorcite{key1} \yearcite{key1}, \yearcite{key2}).     
They also notice that not only thermal Comptonization but also bulk Comptonization by wind motion    
plays a crucial role in spectral formation.    
Quite recently \citet{Narayan} solved the GR radiative transfer problem by post-processing   
GR radiation magnetohydrodynamic (RMHD) simulations and obtained similar results.    
The presence of massive outflow was also pointed out from the observational points of view   
by \citet{key010} and \citet{key300}.       
       
We here make more extensive study of Comptonized spectra expected from super-Eddington       
accretion flow for a variety of black hole masses and see how the spectra depend on       
the black hole mass and the viewing angle.       
So far the theoretical efforts have been focused on clarifying how the spectra depend        
on mass accretion rate,    
while the black hole mass dependence of that flows has been poorly investigated.    
In addition, we pay particular attention to the detailed spectral shape in the hard X-ray ranges.    
Now is a good time to perform such study, since thanks to NuSTAR we now have rich data of    
hard X-ray emission spectra of ULXs with good resolution.       
The organization of the paper is as follows: we first explain our model and methods       
of calculations in section 2 and then show our results in section 3.       
The final section is devoted to discussion and conclusions.

       
\section{Numerical Method}       
\subsection{Radiation Hydrodynamic Simulations of Supercritical Accretion Flows}       
In the present study, we perform spectral calculations      
based on the two-dimensional (2D) radiation hydrodynanmic (RHD) simulation data,      
in which both of bulk and thermal Comptonization are taken into account \citep{key51}.      
This 2D RHD code solves the axisymmetric two-dimensional radiation hydrodynamic equations     
in the spherical coordinates.       
The flux-limited diffusion approximation is adopted (\cite{key3}; \cite{key7}),       
and general relativistic effects are incorporated by adopting the pseudo-Newtonian potential       
\citep{key5}. We also adopt the $\alpha$ viscosity prescription \citep{key6}       
and we set $\alpha=0.1$.       
    
Basic equations and numerical methods are the same as those in       
\authorcite{key1} (\yearcite{key1}, \yearcite{key2}).       
Simulation settings are roughly the same as those in \citet{key2}.      
The computational domain of the radiation hydrodynamic simulations is described by    
$r_{\rm in}=2r_{\rm s}\leq r \leq r_{\rm out}=1000r_{\rm s}$,    
and $0 \leq \theta \leq\pi/2$.     
Here $r_{\rm s}=2GM_{\rm BH}/c^{2}$ is the Schwarzschild radius     
with $G$ being the gravitational constant.    
Grid points are distributed according to the radial and azimuthal coordinates      
and each grid spacing is $\triangle\log_{10} r = (\log_{10} r_{\rm out}-\log_{10} r_{\rm in})/N_{r}$       
and $\triangle\cos \theta=1/N_{\theta}$, respectively, where the numbers of grid points     
are taken to be $(N_{r},N_{\theta})=(192,192)$ throughout the present study.    
    
We start calculations with an empty space but for numerical reasons      
we initially put hot, rarefied, and an optically thin atmosphere.       
Mass is injected continuously with a constant rate of      
$\dot{M}_{\rm input}=10^{3}L_{\rm Edd}/c^{2}$       
through the outer boundary at $r=r_{\rm out}$ and $0.45\pi\leq\theta\leq0.5\pi$.      
We set that the injected matter has an angular momentum corresponding to the       
Keplerian angular momentum at $r=300r_{\rm s}$.       
The matter can go out freely through the boundary $r=r_{\rm out}, 0\leq\theta\leq0.45\pi$,       
and the absorbing boundary condition is adopted at $r=r_{\rm in}$.       
       
The black hole mass is a free parameter and is set to be       
$M_{\rm BH}=10^{1}M_{\odot},10^{2}M_{\odot},10^{3}M_{\odot}$, and $10^{4}M_{\odot}$.      
(The previous study by \cite{key2} solved only the case with $M_{\rm BH}=10M_{\odot}$.)

\subsection{Monte Carlo Calculations of Radiative Transfer}       
The code which we use for the spectral calculation is that developed by \citet{key2}.      
This three dimensional Monte Carlo simulation code (hereafter 3D MC code) solves       
the three-dimensional radiative transfer by means of the Monte Carlo method       
by post-processing the simulation data produced by the 2D RHD code described above.     
Before spectral calculations we time average gas mass density, temperature, and velocity       
at each grid point for every $0.25 M_{\rm BH}/M_{\odot}$ sec.       
We only use the data after accretion flows become quasi-steady.      
          
The 2D RHD simulations data given in terms of the spherical coordinates  
are converted 
to the data in the Cartesian coordinates by interpolation.      
The computational domain of the radiative transfer calculations is set by  
$-300r_{\rm s}\leq x,y,z \leq 300r_{\rm s}$  
and the numbers of the grid points are $(N_{x},N_{y},N_{z})=(160,160,160)$.      
In the quasi-steady accretion regime, the accretion rate onto the central black hole,   
which we calculate by summing up mass passing through the inner boundary at $r=r_{\rm in}$,   
is approximately $\dot{M}\sim 200L_{\rm Edd}/c^{2}$.   
Since the mass input rate through the outer boundary is $10^3 L_{\rm Edd}/c^2$, about 80\% of input material is wandering or being drawn away out of the computational domain. 
   
The photosphere is set up  
$\tau_{\rm eff}(\nu)=\sqrt{3\tau_{\rm a}(\nu)(\tau_{\rm a}(\nu)+\tau_{\rm e})}=10$    
calculated along $z$-axis direction from $z_{\rm max}=300r_{\rm s}$ to $-z_{\rm max}$   
(with $\tau_{\rm a}$ and $\tau_{\rm e}$ being the absorption optical depth   
and the electron scattering optical depth, respectively).   
Here, we only consider bremsstrahlung absorption for evaluating $\tau_{\rm a}$. 
 
At each frequency bin the seed photons, of which the number is $6\times 10^5$, 
are generated within a $\tau_{\rm eff}(\nu)<10$ region.  
In this region the generation point of each photon is selected such that the generated photon number per unit volume is proportional to the local bremsstrahlung emissivity.  
We consider free-free absorption and bulk \& thermal Comptonization effects to describe  
the interaction between photons and matter.  
The black hole swallows photons which go through $r<2r_{\rm s}$.  
The spectrum frequency setting is $\nu_{\rm min}=10^{14} {\rm Hz}<\nu <\nu_{\rm max}=10^{21} {\rm Hz}$ 
with $\triangle \log_{10} \nu = (\log_{10}\nu_{\rm max}-\log_{10}\nu_{\rm min})/100$).

\section{Results}   
\label{sec3}   
\subsection{Overall flow structure}       
       
\begin{figure*}[t!]       
 \begin{center}       
  \includegraphics[width=160mm,bb=0 0 2800 1400]{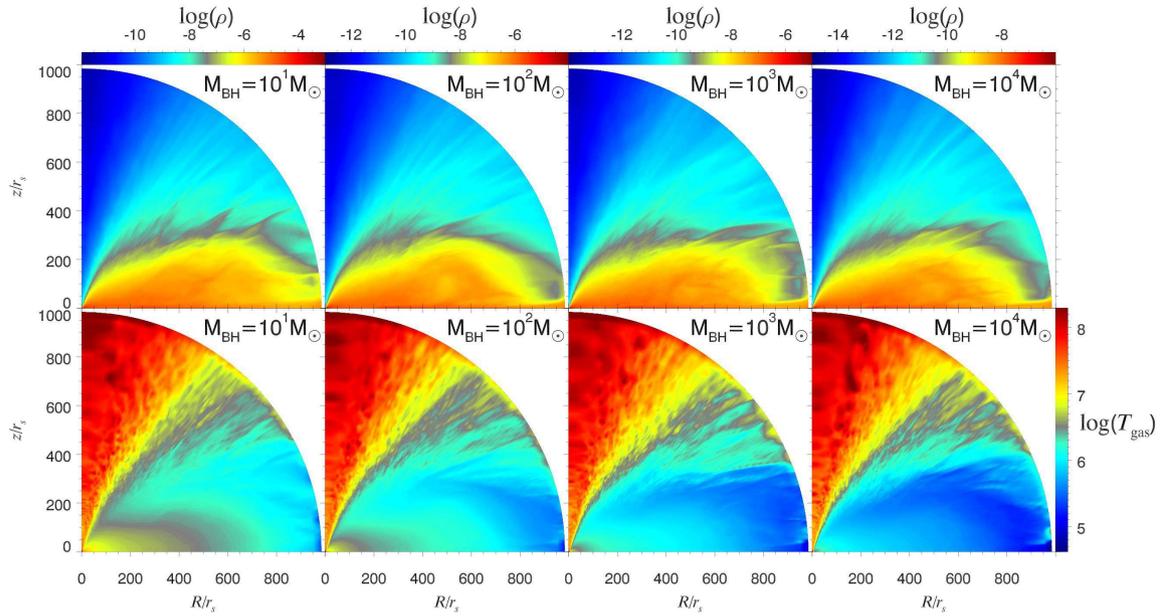}        
 \end{center}       
 \caption{[Top panels]        
Time-averaged density contours of super-Eddington accretion flow        
onto black holes with masses of $M_{\rm BH}=10^{1},10^{2},10^{3}$,and $10^{4}M_{\odot}$        
from the left to the right, respectively. All the panels look similar, if we change   
the color scale according to the density normalization, $\rho \propto M_{\rm BH}^{-1}$.  
 [Bottom panels] The same as those in the upper panels but for the time-averaged gas temperature distributions. The color scales are the same in all the panels.       
 }       
 \label{fig1}       
\end{figure*}       
       
We first overview the flow structure by inspecting the 2D density and temperature     
distributions of each model.      
The top four panels in Figure \ref{fig1} show the time-averaged density contours    
for the black hole mass of $M_{\rm BH}=10^{1},10^{2},10^{3}$, and      
$10^{4}M_{\odot}$ from the left to the right, respectively.       
    
We find a quite similar color contour pattern in every top panel,      
as long as the length scale and the density are normalized by  
$r_{\rm s} (\propto M_{\rm BH})$ and by $\rho (\propto M_{\rm BH}^{-1})$, respectively.  
Common features to all the top panels are that   
the flow consists of a high-density inflow part (or a disk) around the equatorial plane    
(indicated by the red-to-yellow color)    
and a low-density outflow part (or a funnel region)   
around the rotational axis (indicated by the blue color).    
Note that the former disk part can be well modeled by a geometrically thick accretion disk.  
    
The density dependence on $M_{\rm BH}$ can be understood in the following way:      
Let us scale other relevant quantities in terms of the black hole mass; e.g.,       
velocity by free-fall velocity, $v_{\rm ff}=c(r/r_{\rm s})^{-1}\propto M_{\rm BH}^{0}$,      
luminosity by $L_{\rm Edd}\propto M_{\rm BH}$,      
and mass accretion rate by ${\dot M}_{\rm Edd}=L_{\rm Edd}/c^2 \propto M_{\rm BH}$. From   
the continuity equation, we have $r^{2}\rho v_{r} \propto \dot{M}$ ($\propto M_{\rm BH}$).    
We finally obtain $\rho \propto M_{\rm BH}^{-1}$.        
      
\begin{figure}[t!]      
 \begin{center}      
  \includegraphics[width=80mm,bb=0 0 700 700]{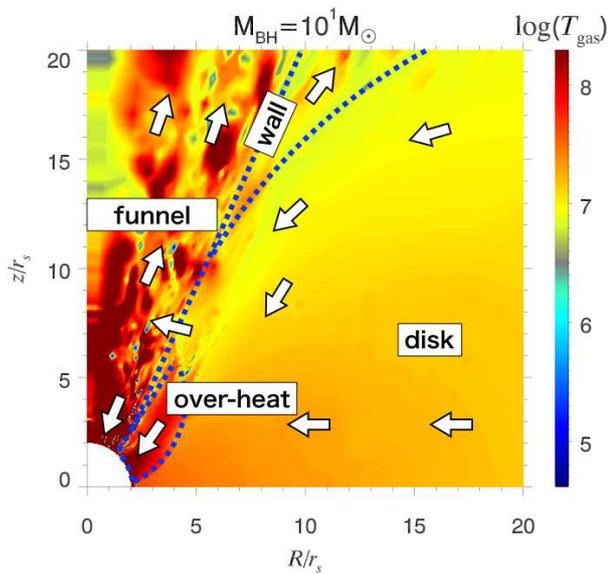}     
 \end{center}      
 \caption{Enlarged snapshot of gas temperature contours close to the black hole at $t=6.04$s    
for model with $M_{\rm BH}=10M_{\odot}$. Overlaid are arrows which represent the direction of    
gas motion. (Note that the lengths of the arrows are not in scale.)     
The three key regions are separated by the blue dash lines:    
accretion flow around the equatorial plane (indicated by {\lq}disk{\rq}),    
funnel region near the polar axis (indicated by {\lq}funnel{\rq}),     
and the over-heated region (indicated by {\lq}over-heat{\rq}).    
The location of the funnel wall is also indicated in this figure (by {\lq}wall{\rq}).    
 }    
 \label{fig2}      
\end{figure}     
    
The bottom four panels in Figure \ref{fig1} show time-averaged gas temperature distribution.      
Again, all the four panels look similar except in the inflow region around the equatorial plane,  
where we find higher temperatures in the left panels than in the right panels.    
This difference can be easily understood, since the temperature of accretion disk is proportional   
to $M_{\rm BH}^{-1/4}$ (detailed explanation will be provided later).    
While the temperature in other regions, especially in the funnel region,  
is insensitive to the black hole mass.     
    
For readers' convenience we specify in figure \ref{fig2}     
the precise locations of the three key regions: disk, funnel, and over-heated region.    
This figure is an enlarged snapshot of the temperature contours    
and corresponds to the bottom left panel of figure \ref{fig1},     
although the panels in figure \ref{fig1} are not snapshots but time-averaged pictures.    
It is important to note that an abrupt heating occurs where the inflow material collides with     
the funnel wall.    
The highest temperature part (with $T\sim 10^{8}$ K) is produced near the black hole     
because of abrupt heating and it is this region plays an essential role in hard photon production    
\citep{key2}.    
    
The disk and funnel regions are distinguished in terms of the directions    
of the gas motion, temperature, and the density (or optical depth).      
Thomson optical depth of the funnel region is:      
\begin{eqnarray}      
  \tau_{\rm e}&=& \int^{z_{\rm max}}_{z_{\rm wall}}\sigma_{\rm T}\rho dz \sim 1.      
\end{eqnarray}      
Here, $\sigma_{\rm T}$ is Thomson cross section, $z_{\rm max}$ is the boundary of computational box, and $z_{\rm wall}$ is the $z$-position of funnel wall.      
The opening angle of funnel is roughly $\theta_{\rm open}\sim 20^{\circ}$--$25^{\circ}$.      

\begin{figure}[t!]      
 \begin{center}      
  \includegraphics[width=80mm,bb=50 50 410 302]{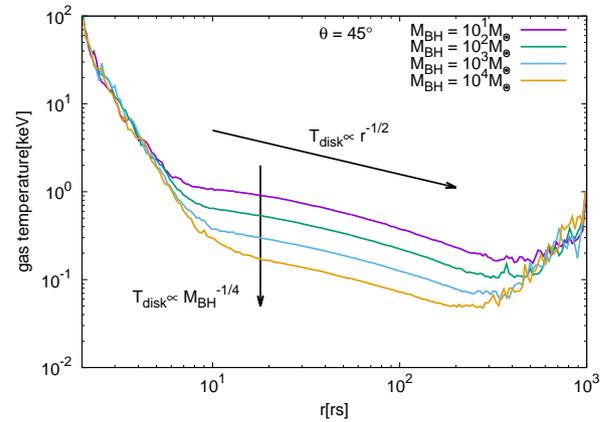}     
 \end{center}      
 \caption{The gas temperature depending on radius ($\theta=45^{\circ}$ from z-axis) for models with $M_{\rm BH}=10^{1},10^{2},10^{3},10^{4}M_{\odot}$.  
 }    
 \label{fig3}      
\end{figure}  

Different mass dependences of the two regions can be understood as follows:      
To demonstrate that the disk temperature at a fixed radius does have a mass dependence of $T_{\rm disk} \propto M_{\rm BH}^{-1/4}$,
we have checked the radial profiles of the disk temperature by using the simulation data (e.g. figure\ref{fig3}),
confirming the relationship of $T_{\rm disk} \propto M_{\rm BH}^{-1/4} \times (r/r_{\rm s})^{-1/2}$.
This mass dependence is the same as that of the standard disk, while the radial dependence is not.
We can simply understand this results in the following way (\cite{key04}, equation 10.22):
The surface temperature of a standard-type disk at a fixed $r/r_{\rm s}$ obeys
\begin{eqnarray}
  \sigma T_{\rm disk}^4 &\sim& \frac{3}{8\pi}\frac{GM\dot{M}}{r^3} \propto M_{\rm BH}^{-1} \left(\frac{r}{r_{\rm s}}\right)^{-3}, \label{eq3}
\end{eqnarray}
leading to $T_{\rm disk} \propto M_{\rm BH}^{-1/4}(r/r_{\rm s})^{-3/4}$, as long as we assume ${\dot M}\propto M_{\rm BH}$. 
In the supercritical flow, by contrast, the temperature profile of the disk no longer depends 
on the mass accretion rate. Then, the disk temperature at a fixed $r/r_{\rm s}$ obeys
\begin{eqnarray}
  \sigma T_{\rm disk}^4 &\sim& \frac{L_{\rm Edd}}{2\pi r^2} \propto M_{\rm BH}^{-1} \left(\frac{r}{r_{\rm s}}\right)^{-2},
\end{eqnarray}
leading to $T_{\rm disk} \propto M_{\rm BH}^{-1/4} (r/r_{\rm s})^{-1/2}$.
Therefore, the supercritical disk has the same mass dependence as that of the standard disk [Eq. (\ref{eq3})].


In the high-temperature funnel region   
($5r_{\rm s}\lesssim r\lesssim 100r_{s}$ and $0^{\circ}\leq\theta\lesssim20^{\circ}$),   
on the other hand, radiative viscous heating balances with Compton cooling.      
That is, by equating   
\begin{eqnarray}    
  \Phi_{\rm vis}&=&\eta \left( r\frac{\partial \Omega}{\partial r}\right)^{2}    
    \sim\alpha E_{0}\Omega_{\rm K}    
      \left(\frac{\Omega}{\Omega_{\rm K}}\right)^{2}    
      \left(\frac{\partial \ln \Omega}{\partial \ln r} \right)^{2},       
\end{eqnarray}       
and      
\begin{eqnarray}       
  \Gamma_{\rm comp}&\sim&  
   4\sigma_{\rm T}c \frac{k_{\rm B}T_{\rm funnel}}{m_{\rm e}c^{2}}  
      \left(\frac{\rho}{m_{\rm p}}\right)E_{0},       
\end{eqnarray}       
we have       
\begin{eqnarray}       
  k_{\rm B}T_{\rm funnel}&\sim&      
     \frac{\alpha}{2}m_{\rm e}c^{2}      
     \frac{m_{\rm p}}{\rho\sigma_{\rm T}r_{\rm s}}      
     \left(\frac{v_{\phi}}{c}\right)^{2}\left(\frac{r}{r_{\rm s}}\right)^{-\frac{1}{2}}      
     \left(\frac{\partial \ln \Omega}{\partial \ln r} \right)^{2}.       
\label{Tfunnel}     
\end{eqnarray}       
We hence see that      
the right-hand side has no mass dependence, so does the funnel temperature.     
     
Numerically, we find $\sim 10^7$ K from equation (\ref{Tfunnel})    
and this temperature is in good agreement with that of the funnel wall     
located along the line of $\sim 20^\circ$ from the rotation axis (see figure \ref{fig2}).     
The gas temperature within the funnel is by a factor of several to ten times greater     
than the above estimation. This is probably because the gas within the funnel      
had been heated when passing through the over-heated region. We should  
also point out that the heating process within the funnel may not be so accurately   
described by the $\alpha$ viscosity prescription.   
This point will be improved by future MHD simulations.  
   
The temperature in the over-heated region is about $T_{\rm heat} \sim 10^8$ K     
regardless of the black hole mass (see figure \ref{fig2} and also \ref{fig1}),     
as is expected from the relation that kinetic energy is converted to internal energy;     
that is, from     
\begin{equation}       
  \frac{1}{2}\rho v^{2} \propto(\gamma -1)\frac{\rho k_{\rm B}T_{\rm heat}}{\mu m_{\rm p}},       
\end{equation}      
we have      
\begin{equation}      
  k_{\rm B}T_{\rm heat}\propto \frac{\mu m_{\rm p}v^{2}}{2(\gamma -1)} \propto M_{\rm BH}^0,      
\end{equation}       
Note that the inflow velocity $v$ is independent black hole mass.     
In reality the temperature of the over-heated region is much lower than     
this simple estimation because of significant Compton cooling by soft photons     
emerging from the underlying accretion flow (with temperature of $T_{\rm disk} \sim 10^7$ K for $M_{\rm BH}=10M_{\odot}$ ).     
     
\subsection{Emergent radiation spectra}       
       
\begin{figure}[ht]       
 \begin{center}       
   \includegraphics[width=80mm,bb=50 50 410 302]{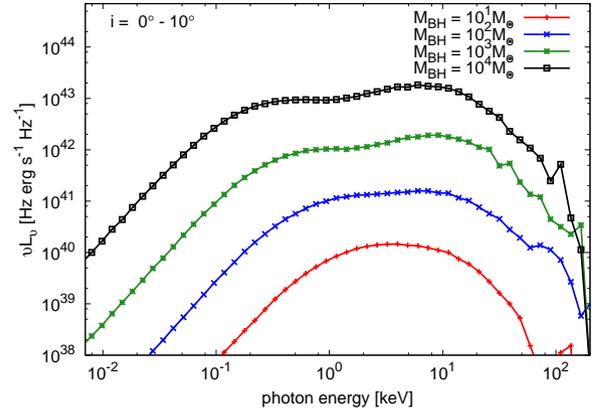}        
 \end{center}        
 \caption{Calculated spectra viewed from the angle of ${\rm i}=0^{\circ} - 10^{\circ}$ from the polar angle 
 for the black hole masses of $M_{\rm BH}=10^{1},10^{2},10^{3}$, and $10^{4}M_{\odot}$.       
 The mass accretion rate is fixed to be ${\dot M}=10^3 L_{\rm Edd}/c^2$.
 The non-monotonic variations in the hard X-ray ($\gtrsim 10$ keV) are
 due to noise arising from insufficient number of photons used in the Monte Carlo methods.
 }       
 \label{fig4}       
\end{figure}      
      
\begin{figure*}[t!]       
 \begin{center}       
   \includegraphics[width=160mm,bb=50 50 1490 626]{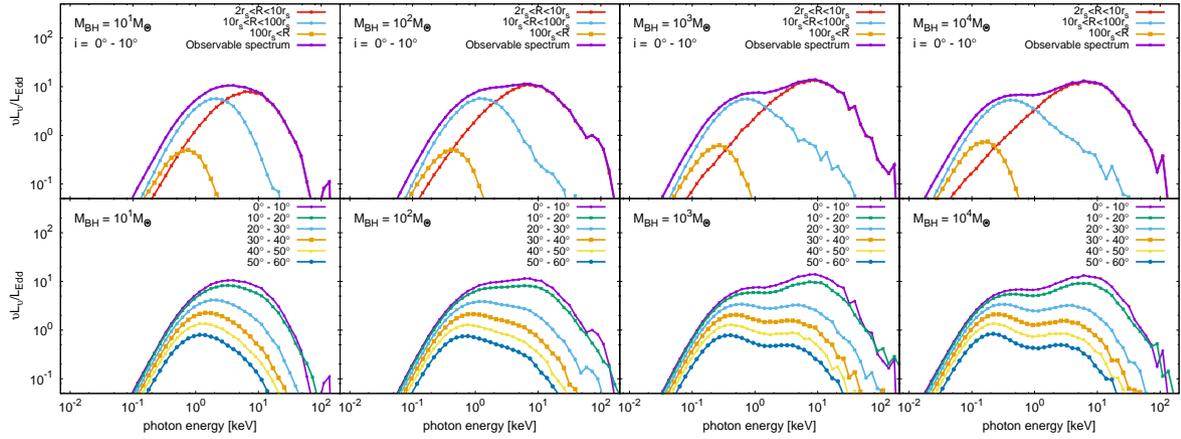}        
 \end{center}       
 \caption{       
   [Top panels] Total emission spectra viewed from the angle of ${\rm i} = 0^{\circ}- 10^{\circ}$      
   (the purple line) and the contributions to the total spectra by different zones:        
   by the inner region between $2r_{\rm s}<R<10r_{\rm s}$(the red line),        
   by the middle region between $10r_{\rm s}<R<100r_{\rm s}$ (the blue line),        
   and by the outer region between $R>100r_{\rm s}$ (the yellow line)        
   for the black hole mass of $M_{\rm BH}=10^{1},10^{2},10^{3}$, and $10^{4}M_{\odot}$,       
   from the left to the right, respectively.       
   [Bottom panels]        
   The same as the top panels but show the viewing angle dependence of the spectra.
   The non-monotonic variations in the hard X-ray ($\gtrsim 10$ keV) are
   due to noise arising from insufficient number of photons used in the Monte Carlo methods.
 }       
 \label{fig5}       
\end{figure*}       
       
Figure \ref{fig4} shows total spectra viewed by a distant face-on observer and its dependence     
on the black hole mass: $M_{\rm BH}=10^{1},10^{2},10^{3}$, and $10^{4}M_{\odot}$.      
Here, by a face-observer we mean an observer viewing from the angle between      
${\rm i}=0^{\circ} - 10^{\circ}$ from the rotation axis.        
    
There are some noteworthy features seen in figure \ref{fig4}.       
First, the larger black hole mass is, the brighter becomes the flow.        
This is because we fixed the mass accretion rate in terms of $M_{\rm BH}$ and, hence,  
the luminosity of super-Eddington accretion flows is roughly proportional   
to the Eddington luminosity, $\nu L_{\nu}\sim L_{\rm Edd}\propto M_{\rm BH}$.  
     
Second, the overall spectral shape in the hard X-ray bands looks quite similar,       
if we normalize the radiation intensity by $M_{\rm BH}$.        
In other words, the peak frequency of the hard component remains the same.       
This is a direct consequence that the temperature of the hard X-ray emitting regions       
is insensitive to the black hole mass; $T_{\rm funnel}\propto M_{\rm BH}^{0}$  
(see the previous subsection).   
The spectral rollover at $h\nu_{\rm peak}\sim 7$ keV can be understood,   
if the photons from the disk region is once over-heated and is then     
Compton cooled by the gas at the funnel wall (with $\sim 10^7$ K).      
     
Finally, the soft X-ray spectra have a mass dependence. That is,       
the higher the black hole mass is,  
the higher becomes the soft X-ray luminosity, and  
the lower becomes the spectral break between the Rayleigh-Jeans part  
and the flat spectrum part       
($h\nu_{\rm break}\propto k_{\rm B}T_{\rm disk}\propto M_{\rm BH}^{-1/4}$).             
The luminosity in the Rayleigh-Jeans range obeys $\nu L_{\nu}\propto M_{\rm BH}^{7/4}$,    
since from   
\begin{eqnarray}       
  T(R)&\equiv&T_{0}\left(\frac{R}{r_{\rm s}}\right)^{-p} \quad{\rm with~} T_{0}\propto M_{\rm BH}^{-1/4},       
\end{eqnarray}       
we find    
\begin{eqnarray}       
  F_{\nu}&\propto&\int_{R_{\rm in}}^{R_{\rm out}} 2\pi R B_{\nu}dR\\       
  &=&\frac{4\pi \nu^{2}k_{\rm B}T_{0}r_{\rm s}^{2}}{c^{2}}  
    \int_{R_{\rm in}/r_{\rm s}}^{R_{\rm out}/r_{\rm s}}  
     \left(\frac{R}{r_{\rm s}}\right)^{1-p} d\left(\frac{R}{r_{\rm s}}\right)\\       
  &\propto&T_{0}r_{\rm s}^{2} \propto M_{\rm BH}^{7/4}    
  ~~~~~~~~~~({\rm for}~ h\nu \ll k_{\rm B}T_{0})       
\end{eqnarray}       
(Here, $R_{\rm in}$ and $R_{\rm out}$ are the radii of the inner and outer edges of the accretion disk, respectively. and we assumed that both radii are scaled in terms of $r_{\rm s}$.)   
This dependence is in good agreement with the simulation results shown in figure \ref{fig4}.   
For these reasons the spectrum is roughly the summation of  
the high- and low-temperature blackbody emissions.   
       
Figure \ref{fig5} displays decomposition of the total emission spectra according to the regions      
where photons are originally generated.      
Here we divide the total flow region into the three:       
the inner region ($2r_{\rm s}<R<10r_{\rm s}$),        
the middle region ($10r_{\rm s}<R<100r_{\rm s}$),        
and the outer region ($R>100r_{\rm s}$).       
Here, $R$ is the radial coordinate in the cylindrical coordinates $R=\sqrt{x^{2}+y^{2}}$.        
      
 From this figure we understand that the observable spectrum is mainly composed by       
direct soft photons from accretion disk        
and hard photons which are Compton up-scattered near the black hole.       
But we see a hard X-ray rollover at around $7$ keV        
which arises due to the Compton down-scattering in low temperature outflow \citep{key2}.       
This issue (regarding the origin of hard X-rays) will be discussed in more details in the next section.    
       
The bottom four panels in Figure \ref{fig5} show the viewing-angle dependence        
of the emergent spectra for each model.       
The larger the viewing angle is, the weaker the hard X-ray flux becomes.       
This is because an observer viewing from large angles is hard to see directly funnel near the black hole ($z\lesssim 30r_{\rm s}$).
We can thus explain the very soft X-ray spectra of ULSs (Ultra-Luminous Supersoft sources;    
see \cite{key0100}, \cite{key33}). 

\section{Discussion}       
\subsection{Brief summary}     
In the present study we elucidate the spectral properties of the super-Eddington accretion     
flow and outflow by means of the three dimensional Monte Carlo radiation transfer simulation based   
on the global 2D RHD simulation data.     
The purpose of the present study is two-fold:    
(1) to extend the study by \citet{key2} 
for a variety of black hole masses, and (2) to calculate more accurate hard X-ray spectra.     
Regarding the first issue, we have seen remarkably similar flow structure and     
overall spectral properties, especially in the hard energy bands.    
The detailed inspection of the hard spectral component (the second issue)    
is made possible by increasing photons used in the Monte Carlo simulations.     
We find a significant excess over the exponential rollover above several keV.     
This should be a result of complex, multiple Compton scattering of photons     
within the over-heated region, as well as in the funnel region.  
We admit the limitations of the present RHD simulations,  
since MHD processes are not properly solved there.  
Definitely, we need radiation-MHD simulation in future work to calculate  
more precisely flow temperature and radiation spectra.  
  
In the following subsections we make more detailed study how photons finally acquire high energy      
by multiple Compton scatterings, interacting with high-temperature or high-velocity plasmas.     
       
\subsection{How is the power-law component constructed?}      
     
\begin{table}    
\tbl{}{    
\begin{tabular}{lll}     
\hline     
 place & quantities (unit) & numerical values \\     
\hline     
photon generation & photon energy (keV) & $1.0 (+1.2/-1.0)$ \\     
& $R$-coordinate ($r_{\rm s}$) & $4.6 \pm 2.6$ \\     
& $z$-coordinate ($r_{\rm s}$) & $2.0 \pm 1.4$ \\      
maximum energy & photon energy (keV) & $100 \pm 85$ \\  
& gas temperature (keV) & $13 \pm 9.2$ \\    
& gas velocity ($c$)  & $0.28 \pm 0.07$ \\    
& $R$-coordinate ($r_{\rm s}$) & $2.2 \pm 2.2$ \\     
& $z$-coordinate ($r_{\rm s}$) & $3.4 (+5.8/-3.4)$ \\    
last scattering & energy amplification$^{*1}$ & $1.1 \pm 0.34$ \\     
& gas temperature (keV) & $1.7 \pm 0.65$ \\      
& gas velocity   ($c$) & $0.19 \pm 0.11$ \\     
& $R$-coordinate ($r_{\rm s}$) & $11 (+16/-11)$ \\     
& $z$-coordinate ($r_{\rm s}$) & $32 (+41-32)$  \\     
\hline    
\end{tabular}} \label{tab1}     
\begin{tabnote}    
Statistical properties of hard photons (with final energy above 10 keV) which reach an observer viewing       
from the angles between ${\rm i}=0^{\circ}$--$10^{\circ}$ for the case with $M_{\rm BH}=10M_{\odot}$.    
Generated soft photons are once heated up to several tens of keV at maximum near the black hole     
and are then Compton cooled down to be observed.  
$^{*1}$ Ratio of the photon energy before to that after the last scattering.  
\end{tabnote}  
\end{table}     
    
\begin{figure}[t!]       
 \begin{center}       
  \includegraphics[width=80mm,bb=0 0 500 500]{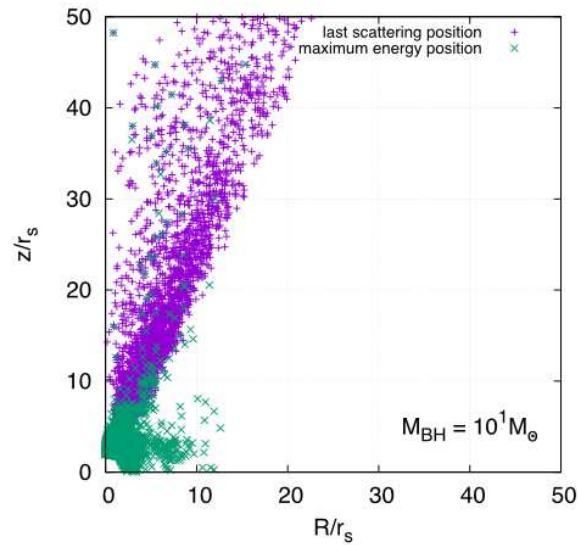}        
 \end{center}     
 \caption{The positions where each hard photon (with final energy above 10 keV)     
suffers a last scattering before reaching a distant face-on observer     
(by the purple pluses ‘+’)     
and those where the same photons reach their maximum energy before the last scattering      
(by the green crosses ‘x’).     
Here, we consider the case with the black hole mass of $M_{\rm BH}=10M_{\odot}$.     
}     
 \label{fig6}       
\end{figure}     
     
\begin{figure}[t!]       
 \begin{center}       
  \includegraphics[width=80mm,bb=0 0 1300 1300]{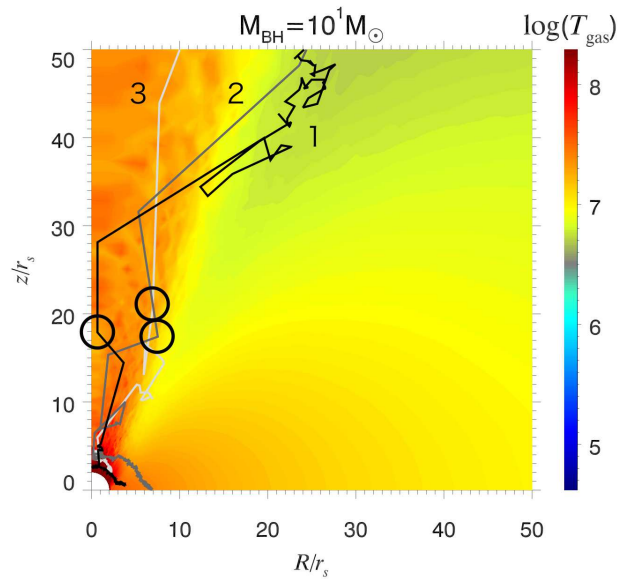}        
 \end{center}       
 \caption{Trajectories of the three representative photons on the $(R, z)$ plane     
overlaid with gas temperature contours.     
The indices 1, 2, and 3 correspond to those in Figure \ref{fig8}.   
Photons escape to observer on the viewing angle ${\rm i}=0^{\circ}$--$10^{\circ}$ at black circle.     
}     
 \label{fig7}       
\end{figure}     
     
\begin{figure}[t!]       
 \begin{center}       
   \includegraphics[width=80mm,bb=50 50 410 698]{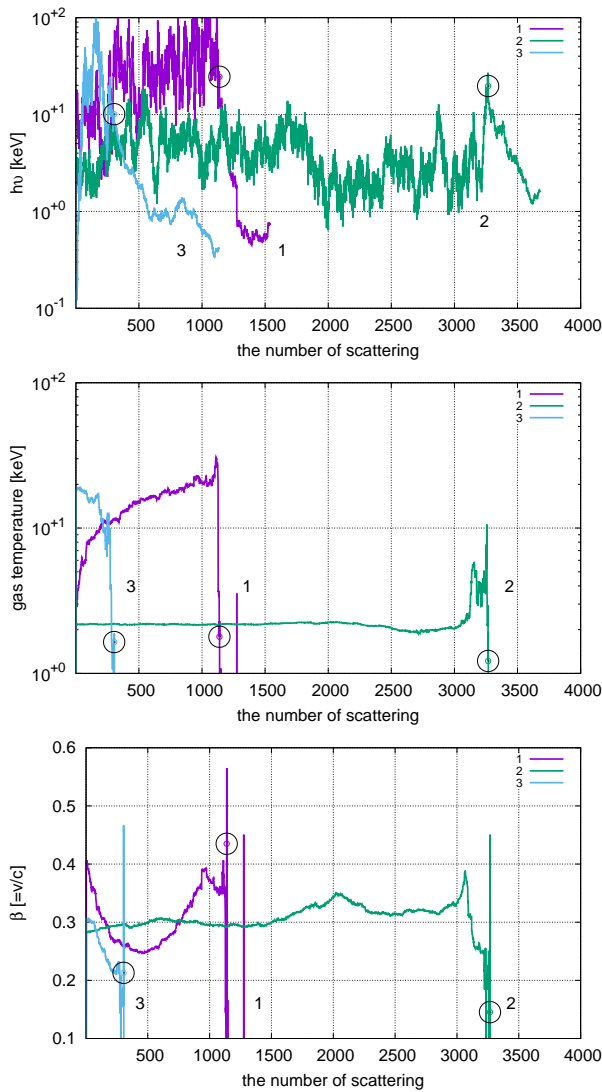}        
 \end{center}       
 \caption{Photon energy, gas temperature, and gas velocity variations which the three photons      
displayed in figure \ref{fig7} experience as functions of the number of scattering.      
Note that the indices,1, 2, and 3 correspond to those in Figure \ref{fig7}.     
Photons escape to observer on the viewing angle ${\rm i}=0^{\circ}$--$10^{\circ}$ at black circle.     
}     
 \label{fig8}       
\end{figure}     
     
In figure \ref{fig4} we have already seen the excess component      
in the hard X-ray energy range (above several keV).      
One possible origin for creating power-law photons is unsaturated Comptonization      
(e.g. \cite{key510}), 
since we estimate the electron scattering optical depth     
to be $\tau_{\rm es}\gg 1$ because of very high density, whereas     
the Compton $y$-parameter [$y\equiv(4k_{\rm B}T/m_{\rm e}c^2)\max(\tau_{\rm es},\tau_{\rm es}^2$)]     
moderately exceeds unity (around several to 10).      
     
In order to understand the origin of hard photons     
we check the trajectories and energy variation histories of the hard photons. 
We set two conditions to select hard photons; (1) their final energy should exceed 10 keV  
and (2) they should eventually be observed by a distant face-on observer.     
    
The results are summarized in Figures \ref{fig6} -- \ref{fig8} for the case with  
$M_{\rm BH} = 10 M_\odot$ (see also Table 1). From these we confirm the following facts:      
\begin{enumerate}     
\item These hard photons were generated as soft photons with $\sim 1$ keV     
within the inflow region around the equatorial plane; $R\lesssim 10r_{s}$ and $z\lesssim 5r_{s}$.      
\item After the generation these photons travel around the black hole for a while, 
being scattered many, many times. 
\item The photons then go into the over-heated region (see figure \ref{fig2} in section \ref{sec3}).      
In this region, optical depth and electron temperature are $\tau_{\rm e}\approx 10$    
and $k_{\rm B}T_{\rm e}\approx 8.6$ keV, respectively, while 
the Compton $y$ parameter is very large; $y \approx 7$.       
The photons are thus Compton up-scattering many times within the over-heated region    
so that they can acquire large energy to become hard photons.    
\item These hard photons then eventually reach the foot-point of the funnel and     
enter the funnel region, where gas is accelerated upward by radiation-pressure force and thus     
has a large radial velocity, up to $\beta=v/c\sim 0.2$, near the funnel wall.     
\item Within the funnel the hard photons experience Compton up- and down-scatterings.  
After passing through the funnel wall, some photons escape to directly reach the observer,  
while some others are reflected by the wall, return to the funnel region and are 
Compton down-scattered there again,  
and finally goes out of the funnel region to reach the observer as soft photons.  
\end{enumerate}        
To summarize, the hard photon production processes are too complex to simply describe.  
The observed hard power-law spectra are formed as a consequence of such complex  
matter-photon interactions.   
       
As \citet{key2} claimed, not only the thermal Comptonizaion but also the bulk Comptonization 
play important roles in the formation of hard X-ray spectra (see their fig. 5).     
The speed of the outflow that photons encounter is around $\beta\equiv v/c\sim 0.2$    
(see table \ref{tab1}). Since the photon energy amplification factor is from \citet{key51} 
\begin{eqnarray}       
  \frac{\triangle (h\nu)}{h\nu}    
     &=&\frac{\frac{4}{3}\beta^{2}\gamma^{2}m_{\rm e}c^{2}-h\nu}{m_{\rm e}c^{2}}.     
\end{eqnarray}      
we estimate that the photon energy can acquire energy at most around     
$h\nu\sim \frac{4}{3}\beta^{2}\gamma^{2}m_{\rm e}c^{2}\sim 30 (\beta/0.2)^2$ keV.     
For $\beta=0.1$ -– $0.3$ we find $h\nu \simeq 8 - 70$ keV.     
       

\subsection{The effects of iterated temperature, magnetic field, and general relativity.}
We do not recompute the temperature in their simulations using the Monte Carlo code, although this is sometimes implemented by others using similar methods (e.g. \cite{Narayan}).
Here, we calculate what the {\lq\lq}correct{\rq\rq} temperature would be, if we would use
the calculated radiation spectra as a source of Compton cooling.
For this purpose, we check the temperature of the photosphere 
at around $(R, z)=(11 r_{\rm s}, 32 r_{\rm s})$,
from which hard X-ray ($> 10$keV) photons mainly come from,
finding that the mean photon energy is $<h\nu>\sim 14.9$ keV.
This energy corresponds to the radiation temperature of $k_{\rm B}T_{\rm rad}^{1} = <h\nu>/4 \sim 3.7$ keV.
On the other hand, the gas temperature found in the original RHD simulation
is $k_{\rm B}T_{\rm gas}^{0}  \sim 1.2$ keV. Therefore, we have $T_{\rm gas}^{0} < T_{\rm rad}^{1}$, meaning that 
gas should be Compton heated by radiation. However, we should point that 
bulk Compton is dominant over the thermal Compton even for this recalculated temperature;
i.e., $v_{\rm thermal}/c \sim \sqrt{3k_{\rm B}T_{\rm gas}^{1} / m_{\rm e} c^2} \sim 0.147$ (assuming $k_{\rm B}T_{\rm gas}^{1}\sim k_{\rm B}T_{\rm rad}^{1}$), 
while the bulk velocity is $v_{\rm bulk}/c  \sim 0.2$.
We can thus safely conclude that the iterated temperature can hardly affect the hard X-ray emission properties discussed in the present paper.

We consider why recomputation of the temperature is not so critical in our simulations unlike the simulations by \citet{Narayan}.
\citet{Narayan} mentioned that the gas temperature (before the recomputation) is very high around the rotation axis since the Compton-cooling is less effective.
The cause of that is the deficit of the radiation induced by the M1-closure method.
The M1 method tends to suppress the radiation energy density around the axis via the artificial centrifugal shocks (\authorcite{key511} \yearcite{key511}; \yearcite{key522})
Thus, after recomputation, the high gas temperature ($\sim 10^{9}$ K) is significantly reduced to about $10^{8}$ K by the effective Compton cooling.
In contrast, artificial reduction of the radiation energy density around the axis does not occur in the FLD approximation method that is employed in the present work.
Hence, the gas temperature near the polar axis may be already low, about $10^8$ K.
In addition, the relatively low gas temperature in the region far from the polar axis is nearly the same as those reported by \citet{Narayan}.
In this region, the gas temperature does not change so much by the recomputaion.
For these reason, our spectra look quite similar to those in \citet{Narayan}.

The differences in methods (MHD versus non-MHD, general relativity versus no general relativity) are surprisingly small.
To check if this is true, we first estimate the effects of the gravitational redshift.
The spectrum is reproduced by multiplying the redshift to the photon energy at only last scattering position.
The redshift is very important near the black hole, but in that place, photons are many Compton scattered.
Then it seems to be sufficient to consider the redshift at the last scattering position.
There is no significant difference between spectrum, e.g. ${\rm sed(no~redshift)/sed(redshift)} \sim 0.98$ -- $1.2$ each photon energy.
This is because the location of the last scattering surface of the hard X-ray is quite far from the black hole.
For this reason, we conclude that our spectrum (non-GR) is almost same as \cite{Narayan} (GR).

We next check how important magnetic effects are.
It is important in this context to point out the inequality of
radiative-pressure $>$ gas-pressure $>$ magnetic-pressure in the funnel region,
which was first pointed out by radiation-MHD simulations by \citet{key42}.
The effects of the magnetic field seem to be weak and spectrum is almost same between MHD and non-MHD.
However, magnetic dissipation could be more enhanced above or near the photosphere, rather than in the equatorial region.
It is thus necessary to check if this is the case in future radiation-MHD simulations.

We further estimate the effects of cyclotron emission and absorption as follows:
There are two regions in which magnetic field is relatively strong: 
the funnel region and the region very close to the black hole. 
In the funnel region, first of all, 
the magnetic field strength is $B \sim 10^{5}$ G \citep{key42}.
The gas velocity is up to $v < 0.5c$, and the Lorentz factor is $\gamma \sim 1$.
The typical frequency of cyclotron is $\nu_{g}=eB/(2\pi m_{\rm e} c) \sim 3 \times 10^{11}$ Hz,
$\nu_{\rm crit} =3 \gamma^2 \nu_{g} \sin\alpha/2 \sim \nu_{g} \sin\alpha < \nu_{g}$.
In this frequency $\nu_{\rm crit}$, the cyclotron process does not emit and absorb X-ray.

In the region near the black hole ($R<10r_{\rm s}, z<5r_{\rm s}$), secondly,
the magnetic field is $B\sim 10^{8}$ G \citep{key42}.
In the same way, frequency is $\nu_{g} \sim 3\times 10^{14}$ Hz, and so the cyclotron process does not absorb X-ray, and emit radio (or far-infrared) wave.
The seed photon generated near the black hole in the disk are observed as hard X-ray through the Compton effect.
But cyclotron radio emission does not become this seed photon
because there are a lot of the seed photons by free-free emission ($\sim 1$ keV) from the disk.
For these reasons, our spectrum (non-MHD) looks like that by \cite{Narayan} (MHD).

It is, of course, possible that the flow structure might be somewhat altered, if we would incorporate the MHD and GR effects.
Especially, the effects of MHD and GR would be important in magnetically arrested disks (MADs), and MADs around highly spinning BH, respectively.
(We note that we have studied non-MADs around non-rotating BHs in this paper).
For example, \citet{key30000} carried out 3D GR-RMHD simulations of super-critically accreting MADs
and found that the cyclo-synchrotron cooling is dominant in the jet and corona region.
\citet{Narayan} found that the SEDs of MADs around a highly spinning BH differ from those of non-MADs or MADs around a non-spinning BH,
because the speed of jets in MADs with highly spinning BH is higher than the others and that results in the decrease of the optical depth in the jet and leads high frequency photons emitted from the plasma near the BH to be observable.
It is our future work to calculate radiation spectra based on the GR-RMHD simulations.

\subsection{Application to the ULXs}
\begin{figure}[t!]     
\begin{center}     
\includegraphics[width=80mm,bb=0 0 700 500]{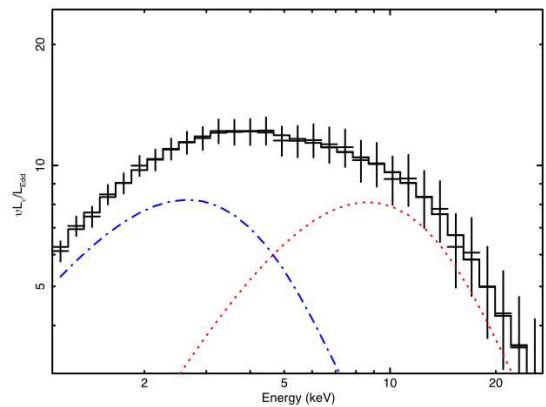}    
\end{center}     
\caption{   
Result of fitting to the theoretical spectrum displayed in figure \ref{fig4}    
for the case with $M_{\rm BH}=10 M_{\odot}$.    
The fitting function is the summation of two blackbody spectra associated with a power-law component    
(${\rm SIMPL*(DISKBB+DISKPBB)}$) in Xspec, same as that adopted in \cite{key8}   
excluding the interstellar absorption (${\rm TBABS}$).   
The red dotted and blue dash-dotted curves, respectively, represent  
the best-fit high-temperature and low-temperature blackbody components.  
[More precisely, the former component is the $p$-free disk blackbody spectra,  
in which the disk surface temperature is scaled as $T_{\rm surface} \propto r^{-p}$  
with $p$ being the fitting parameter; see \cite{key302}.]  
}    
\label{fig9}     
\end{figure}

\begin{table}    
  \tbl{}{    
    \begin{tabular}{lll}     
      \hline     
      Model & Parameter (unit) & numerical values \\     
      \hline     
      ${\rm DISKBB}$ &$T_{\rm in}$ (keV)&$ 1.0(\pm 0.49)$\\   
      ${\rm DISKPBB}$ &$T_{\rm in}$ (keV)&$ 3.1(+1.6/-1.1)$\\   
      &$p$&$0.87(+0.13/-0.27)$\\   
      ${\rm SIMPL}$ &$\Gamma$&$2.8(+2.2/-2.8)$\\   
      \hline    
  \end{tabular}} \label{table2}     
  \begin{tabnote}    
    The fitting parameters obtained by Xspec for the case with $M_{\rm BH}=10M_{\odot}$ (see, Figure \ref{fig9}).   
  \end{tabnote}    
\end{table}     
   
In figure \ref{fig4} we show the observable photon energy spectrum which is to be observed    
by a face-on observer 
for the case with $M=10 M_{\odot}$.    
We find a significant excess over an exponential rollover in the hard X-ray range.    
This is quite reminiscent of the recent NuSTAR observations of ULXs.  For example,    
\citet{key8} reported a hard X-ray excess in the NuSTAR spectrum of Holmberg II X-1,    
one of the most extensively studied ULXs. They performed spectral fitting,    
finding that this excess component can be best-fit by a power-law component with the    
photon index of $\Gamma=3.1^{+0.3}_{-1.2}$.    
They discuss that likely origin of this power-law tail is    
Comptonization of soft photons by a hot (or even non-thermal) coronal plasma.    
   
To check to what extent our theoretical spectra can reproduce the observed ones,    
we perform similar spectral fittings to the theoretical one     
by using the same spectral models excluding ${\rm TBABS}$ (i.e., the model which X-ray is absorbed by interstellar medium) as those used by \citet{key8}.  
The results are illustrated in figure \ref{fig9}. It clearly shows that  
the theoretical spectrum can nicely be represented by two blackbody-like components   
with a power-law tail (see also table 2 for the best-fit parameters).  
The most remarkable is the power-law photon index, $\Gamma\sim 3$, which is 
in good agreement with the NuSTAR observation of Ho II X-1 [$\Gamma = 3.1~(+0.3/-1.2)$].   
The temperature of the higher temperature blackbody is $\sim 3$ keV, somewhat   
higher than the observed one [$1.8~(+0.7/-0.3)$ keV].   
   
The temperature of the lower temperature one ($\sim 1.0$ keV) is, on the other hand,   
significantly higher than the observed one [$0.20~(+0.03/-0.04)$ keV],    
But this disagreement should not be taken seriously, since our computational box is  
not large enough to resolve the accretion flow structure at large radii, from where 
soft X-ray emission originates.   
We thus focus out discussion to the hard X-ray properties from the super-Eddington flow   
in the present study.    
We expect that future larger-box simulations will improve this discrepancy.   
   
We may thus safely conclude that the success in reproducing the observed hard excess spectrum    
provides good support to the super-Eddington scenario for ULXs.    

\subsection{The effects of the photosphere}
\citet{key2} carried out the calculations for the other choices of $\tau_{\rm eff}=1,3,5,10,20$ for the locations of the photosphere,
but confirmed that the spectrum for the photosphere $\tau_{\rm eff}=10$ is essentially the same as that obtained for the deeper photosphere $\tau_{\rm eff} >10$.
We also recalculate the spectra for the case of different photosphere, i.e., $\tau_{\rm eff}=5,20,30$.
The spectrum looks very similar and the fitting results are quite similar;
now the best-fit parameter at $\tau_{\rm eff}=30$ is $\Gamma = 3.1(+1.3/-2.1) $,
in good agreement with the case with $\tau_{\rm eff}=10$ (see table\ref{table2}).
For these reason, it is essentially to set photosphere $\tau_{\rm eff}=10$.


\begin{ack}       
This work is partially supported by JSPS Grant-in-Aid for Scientific Research (C)    
 (17K05383 S. M.; 15K05036 K.O.).     
Numerical computations were mainly carried out on Cray XC30 at Center for Computational Astrophysics, National Astronomical Observatory of Japan.       
This research was also supported by MEXT as“Priority Issue on Post-K computer”    
(Elucidation of the Fundamental Laws and Evolution of the Universe) and JICFuS.    
\end{ack}


\end{document}